# The Nuclear Symmetry Energy at Sub-saturation Densities

W.G. Lynch* and M.B. Tsang[#]

*National Superconducting Cyclotron Laboratory and the Department of Physics and Astronomy*

*Michigan State University, East Lansing, MI 48824 USA*

## Abstract

The density dependence of the nuclear symmetry energy governs important aspects of very neutron rich systems such as heavy nuclei and their collisions, neutron stars and their mergers. Many analyses of experimental data have generated constraints on the symmetry energy and its first derivative at saturation density, $\rho_0 \approx 2.7 \times 10^{14}$ g/cm$^3$. We show that each analysis does not accurately constrain the symmetry energy at $\rho_0$, but rather at a lower density, $\rho_s$, that is most sensitively probed by that analysis. Using published constraints on the symmetry energy and its first derivative at $\rho_0$, we constrain the symmetry energy within a density range of $0.25\rho_0$ to $0.75\rho_0$ that is relevant to the inner crusts of neutron stars and to the neutrino-sphere of core-collapse supernovae. With appropriate data, the symmetry energy can be similarly constrained over a wider density range.

*Electronic address: lynch@nscl.msu.edu  
[#]Electronic address: tsang@nscl.msu.edu

The nuclear Equation of State (EoS) is central to the understanding of matter found in neutron stars and in explosive stellar environments [1-4]. This includes the dynamics in neutron star mergers and core collapse supernovae in which many of the heavy elements are formed [4,5]. The description of such neutron-rich environments requires extrapolating the properties of neutron-rich matter from that of symmetric matter containing equal numbers of neutrons and protons [6-8]. This extrapolation is governed by the nuclear symmetry energy, which can be defined to be the difference between the EoS of neutron matter and that of symmetric matter. In addition to its properties at saturation density, $\rho_0 \approx$ 2.7x10$^{14}$ g/cm$^3$, the symmetry energy must be determined at $\rho/\rho_0 \approx 0.25$ to understand the supernova neutrino sphere [4,9,10], and at $0.5<\rho/\rho_0<0.7$ to predict the crust-core boundary in neutron stars [11] and crustal vibrations in Magnetars [12]. At higher densities of $1<\rho/\rho_0<3$, the symmetry energy largely governs tidal deformabilities in neutron star mergers [5] and the mass vs. radius correlation of neutron stars [2]. Laboratory data are being employed to provide constraints on its density dependence [13-21]. Here, we discuss how make such constraints more quantitative by finding the density to which each analysis is most sensitive and extracting the symmetry energy at that density. This provides constraints on the symmetry energy at $0.25\rho_0 < \rho < 0.75\rho_0$ that are directly comparable to theoretical calculations.

For moderate asymmetries, the symmetry energy can be approximated by $\varepsilon_{sym} = S(\rho)\delta^2$, where $S(\rho)$ describes the density dependence of the symmetry energy. The asymmetry $\delta$ is defined by $\delta = (\rho_n - \rho_p)/\rho$, where $\rho_n$, $\rho_p$ and $\rho = \rho_n + \rho_p$ are the neutron, proton and nucleon number densities, respectively [8]. Under the assumption that the symmetry energy is more easily constrained near saturation density, it is customary to expand $S(\rho)$ around the saturation density, $\rho_0$, yielding

$$S(\rho) = S_0 + \frac{L}{3\rho_0}(\rho - \rho_0) + \frac{K_{sym}}{18\rho_0^2}(\rho - \rho_0)^2 + \cdots \qquad (1)$$

where, $L = 3\rho_0 \left.\frac{dS(\rho)}{d\rho}\right|_{\rho=\rho_0}$ provides the pressure, $P_0 = L\rho_0/3$, of pure neutron matter at saturation density [8]. Some constraints on $S_0$ and $L$ can be provided by the bulk and surface symmetry energies of nuclei and other observables, but the uncertainties in these constraints are significant [7,8,13-21].

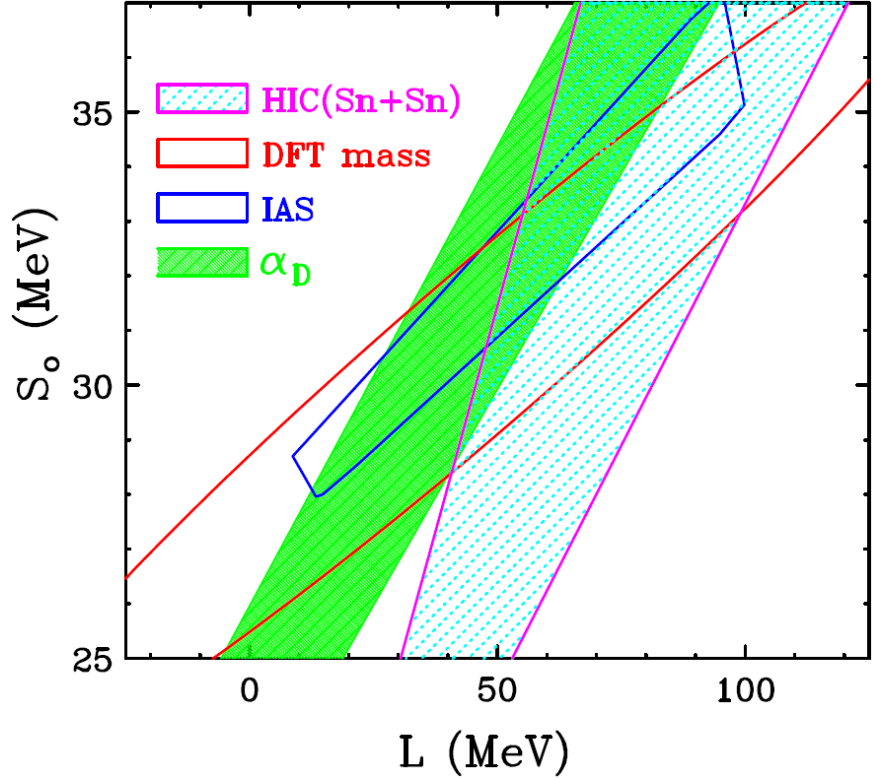

Figure 1 (color online): Contours for the allowed values for ($S_0$, $L$) obtained from analyses of isospin diffusion (labeled HIC Sn+Sn), masses (labeled DFT mass), isobaric analog states (labeled IAS), and electric dipole polarizability of $^{208}$Pb (labeled $\alpha_D$).

Recent constraints on $S_0$ and $L$ from nuclear reaction and structure measurements have been compiled in refs. [7,14,15,16]. The bounded regions in Figure 1 [22] denote correlated constraints on $S_0$ and $L$ from ref. [7] that have been extracted from four different structure and reaction observables. For clarity, this figure shows only one constraint region for each observable, even when others exist [15]. The four bounded regions in Figure 1 correspond to: **1)** analyses of nuclear masses using density functional theory [17] (two red curves as part of an ellipse), **2)** analyses of isobaric analog states [18] (blue contour), **3)** analyses of the electric dipole polarizability for $^{208}$Pb [19, 20, 23] (green shaded region), and **4)** analyses of isospin diffusion measurements in peripheral Sn+Sn collisions at E/A=50 MeV [13, 24, 25] (light blue shaded region).

For each constraint in Figure 1 the relevant experimental observables were modeled with specific functional forms for $S(\rho)$ with internal parameters that were determined by fitting the measured observables. These fitted internal parameter values were used to calculate the corresponding contours of allowed values for $S_0$ and $L$ in Figure 1. Over the years, attempts have been made to reduce these uncertainties in $S_0$ and $L$ by averaging the values of $S_0$ and $L$ from different observables [16] or by considering the overlaps of the various $S_0$ and $L$ constraint contours [15].

Figure 1 shows that the slopes $\frac{\Delta S_0}{\Delta L}$ of the contours obtained via density functional theory for masses (DFT) and isobaric analog states (IAS) are similar, while those obtained from the electric dipole polarizability of $^{208}$Pb and from isospin diffusion (HIC) are larger. In the following, we will show that each slope reflects a specific density $\rho_s$ at which each observable most sensitively probes the symmetry energy. Averaging the best-fit values of $S_0$ and $L$ from different observables, or requiring overlap of constraint boundaries to obtain more stringent constraints on $S_0$ and $L$ could be accurate if all observables were dominated by $S(\rho_s)$ values near $\rho_s \approx \rho_0$. One expects that $\rho_s < \rho_0$ for these observables from the positive correlation between $S_0$ and $L$ and the analyses of refs. [6,7,13,14,18,19,21,28,29]. We will show, however, that the sensitive densities $\rho_s$ lie far below $\rho_0$, requiring extrapolation to $\rho_0$ using analysis dependent functional forms for $S(\rho)$ that differ from one analysis to another. To avoid errors in extrapolation, we therefore determine the symmetry energy $S(\rho_s)$ at the specific density $\rho_s$ for each observable.

To proceed, we recall the analysis of Brown [21] who fit the energies of 11 doubly closed shell nuclei with 18 different Skyrme symmetry energy functionals of the form

$$S(\rho) = a(\rho/\rho_0) + b(\rho/\rho_0)^{1+\sigma} + c(\rho/\rho_0)^{2/3} + d(\rho/\rho_0)^{5/3} \qquad (2)$$

using Hartree-Fock density functional theory [21, 28]. This form of the symmetry energy functional is frequently used to calculate nuclear masses and isobaric analog states. The parameters a, b, c, d and σ fully define the values of $S_0$ and $L$ for this functional [21,28].

The left panel of Figure 2 shows the symmetry energy functionals of Brown. After fitting the masses, these Skyrme functionals intersect near a cross-over point, located at $S(\rho_s) \approx 24.7 \pm 0.8\ MeV$ and $\rho_s/\rho_0 \sim 0.60 - 0.66\ (i.e.\ \rho_s \approx 0.1\ fm^{-3})$ as shown in the left panel of Figure 2. The $S_0$ and $L$ values for these functionals, shown by the three groups of open squares in the right panel of Figure 2, display a linear correlation with a slope (dotted line) that is slightly greater than that of the long axis of the IAS contour (blue). Each group of Brown's calculations corresponds to one of three neutron skin thicknesses for $^{208}$Pb: $R_{np}$ = 0.16, 0.20 and 0.24 fm, with L ≈ 40, 60 and 90 MeV, respectively [21, 29]. Above and below $\rho_s$, $S(\rho)$ diverges into three groups of curves with larger $S_0$ and $L$ values being correlated with larger $R_{np}$.

We note that functionals with a larger (smaller) $L$ display stronger (weaker) density dependencies at $\rho_s$. $L$ can be significantly changed without degrading the agreement with the masses provided that $S_0$ is also modified so that the new value for $S(\rho_s)$ at $\rho_s/\rho_0 \sim 0.63$ is approximately $24.7\ MeV$. This unique fixed point, listed under the cross-over analysis of Table 1, must be satisfied for these functionals to replicate the observed symmetry energy contribution to these nuclear masses. Increasing or decreasing $L$ without correspondingly modifying $S_0$ changes $S(\rho_s)$ and strongly degrades the agreement with experiment.

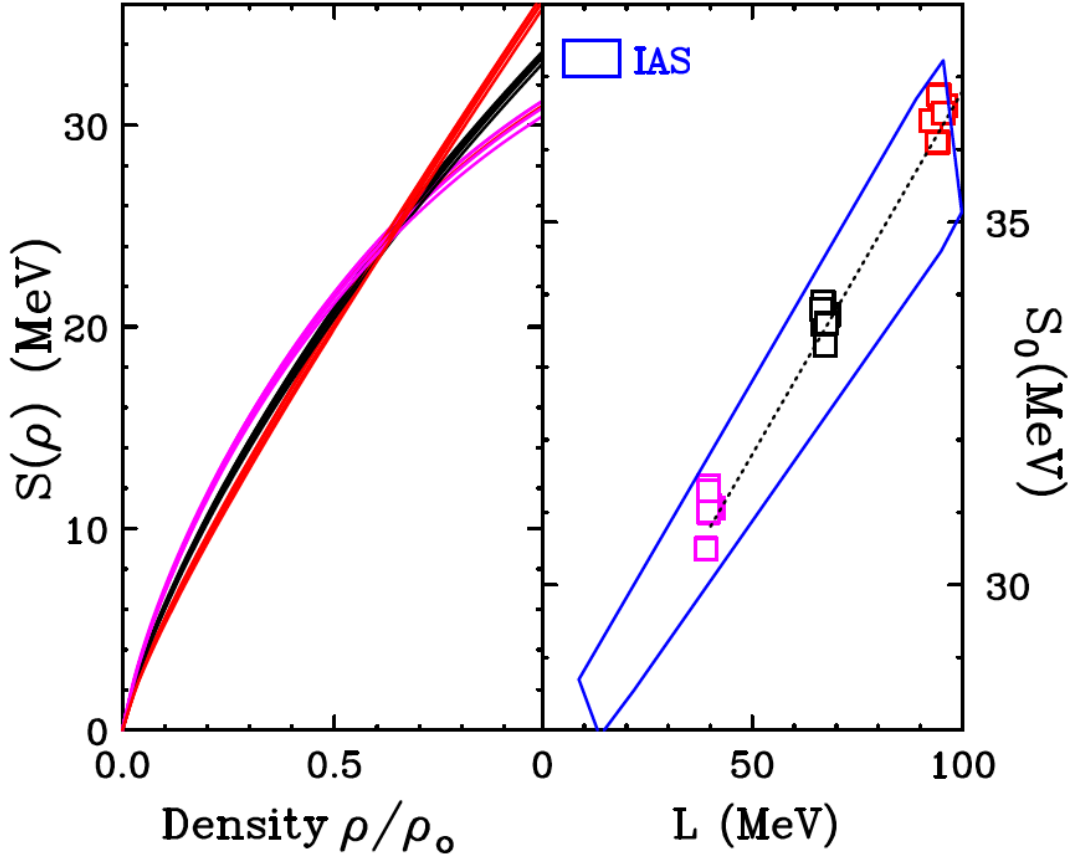

Figure 2: Left panel: Skyrme symmetry energy functions of Eq. (3) best fitted to the masses of double magic nuclei by ref. [21]. The three groups of curves correspond to the open squares in the right panel and predict 3 different values for the "neutron skin" of $^{208}$Pb. Here, $\rho_0 = 0.16\ fm^{-3}$. b) Right panel: The polygon represents the IAS contour also shown in Figure 1. The squares define the dashed guideline and indicate the $S_0$ and $L$ values for the symmetry functions shown in the left panel.

In Brown's analysis, the best-fit values for $S_0$ and $L$ lie along the dashed line in the right panel of Figure 2. Any intermediate value for $S_0$ and $L$ that lies along this line will provide $S(\rho_s) \approx 24.7\ MeV$ and fit the masses nearly as well. Indeed, two such points along this line with a relative displacement of $(\Delta L, \Delta S_0)$ have nearly the same value for $S(\rho_s)$ and provide excellent fits to the masses because the change in $S(\rho_s)$ due to $\Delta L$ (i.e. $\frac{\partial S(\rho_s)}{\partial L}\Delta L$) is canceled by the change in $S(\rho_s)$ due to $\Delta S_0$ (i.e. $\frac{\partial S(\rho_s)}{\partial S_0}\Delta S_0$). In other words, this line lies perpendicular to the gradient of $S(\rho_s)$ in the ($L$, $S_0$) plane and its slope $u(\rho_s)$ is given by

$$u(\rho_s) = \Delta S_0/\Delta L = -\frac{\partial S(\rho_s)}{\partial L}\bigg/\frac{\partial S(\rho_s)}{\partial S_0}. \qquad (3)$$

Calculating this ratio of partial derivatives for any of the symmetry energy functionals discussed in this paper, i.e. Eqs. (2) and (3), reveals $u(\rho_s)$ to be a monotonically decreasing function of $\rho_s$ at $\rho_s < \rho_0$. Setting $u(\rho_s)$ equal to the observed slope $u=0.100\pm0.006$ of the dashed line yields the sensitive density, $\frac{\rho_s}{\rho_0} \approx 0.63 \pm 0.03$ and $S(\rho_s) \approx 24.7 \pm 0.8\ MeV$ after substituting $\rho_s$ into any of Brown's functionals. This constraint, obtained consistently from both the cross-over and slope techniques, is shown as the open square in Figure 3. Constraints in this density region are very relevant to the crust-core boundary in neutron stars [10].

| Method | | Slope analyses | | | Cross-over analyses | |
|---|---|---|---|---|---|---|
| Constraint | u | $\rho_s$ | $S(\rho_s)$ | | $\rho_s$ | $S(\rho_s)$ |
| Brown | 0.100 ±0.006 | 0.63±0.03 | 24.7±0.8 | | 0.63±0.03 | 24.7±0.8 |
| DFT | 0.079±0.002 | 0.72±0.01 | 25.4±1.1 | | | |
| IAS | 0.092±0.008 | 0.66±0.04 | 25.5±1.1 | | | |
| HIC(isodiff) | 0.23±0.06 | 0.24±0.11 | 11.4±1.4 | | 0.24±0.7 | 10.6±1.5 |

Table I: Values of slopes, $\rho_s$ and $S(\rho_s)$ obtained from the direct examination of the symmetry energy cross-over point and from the slope of the correlation for different experimental observables.

The slope technique can be directly applied to the DFT and IAS contours in order to obtain their corresponding sensitive densities and symmetry energies. The DFT constraint in Figure 1 denotes the 2σ contour obtained by fitting the masses of 28 spherical and 44 deformed nuclei and the IAS contour is obtained by fitting Isobaric Analog States with 30≤A≤240. Both analyses utilized Skyrme parameterizations of the form in Eq. (2). The best-fit value for the DFT lies at the center of the 2σ DFT contour. Using the extracted slope $u = 0.079 \pm 0.002$ and the form of the UNEDEF0 symmetry functional used for the DFT contour [30], we extract the sensitive density of $\rho_s/\rho_0 = 0.72 \pm 0.01$ and a symmetry energy $S(\rho_s) = 25.4 \pm 1.1\ MeV$. From the IAS constraint region in Figure 1 with its extracted slope of $0.092 \pm 0.008$ we similarly obtain a sensitive density of $\rho_s/\rho_0 = 0.66 \pm 0.04$ and a symmetry energy $S(\rho_s) = 25.5 \pm 1.1\ MeV$. These values are listed under "slope analyses" in Table 1 and the results are plotted as a solid red circle and blue solid triangle in Figure 3, respectively.

Danielewicz et al. [18,31] have also constrained $S(\rho)$ as a function of density over $0.25 \leq \rho/\rho_0 \leq 1$ by analyzing nuclei in *different* mass ranges *separately* using a wider range of symmetry energy functionals. Figure 3 shows a comparison between the current analyses and 1σ IAS constraint of ref. [18, 31] (blue dashed contour). Our analyses and IAS contour of refs. [18, 31] overlap, both showing that the most sensitive IAS constraint on $S(\rho)$ is at $\rho_s/\rho_0 \approx 0.66 \pm 0.04$ and that $S(\rho_0)$ and L are rather poorly constrained by the data.

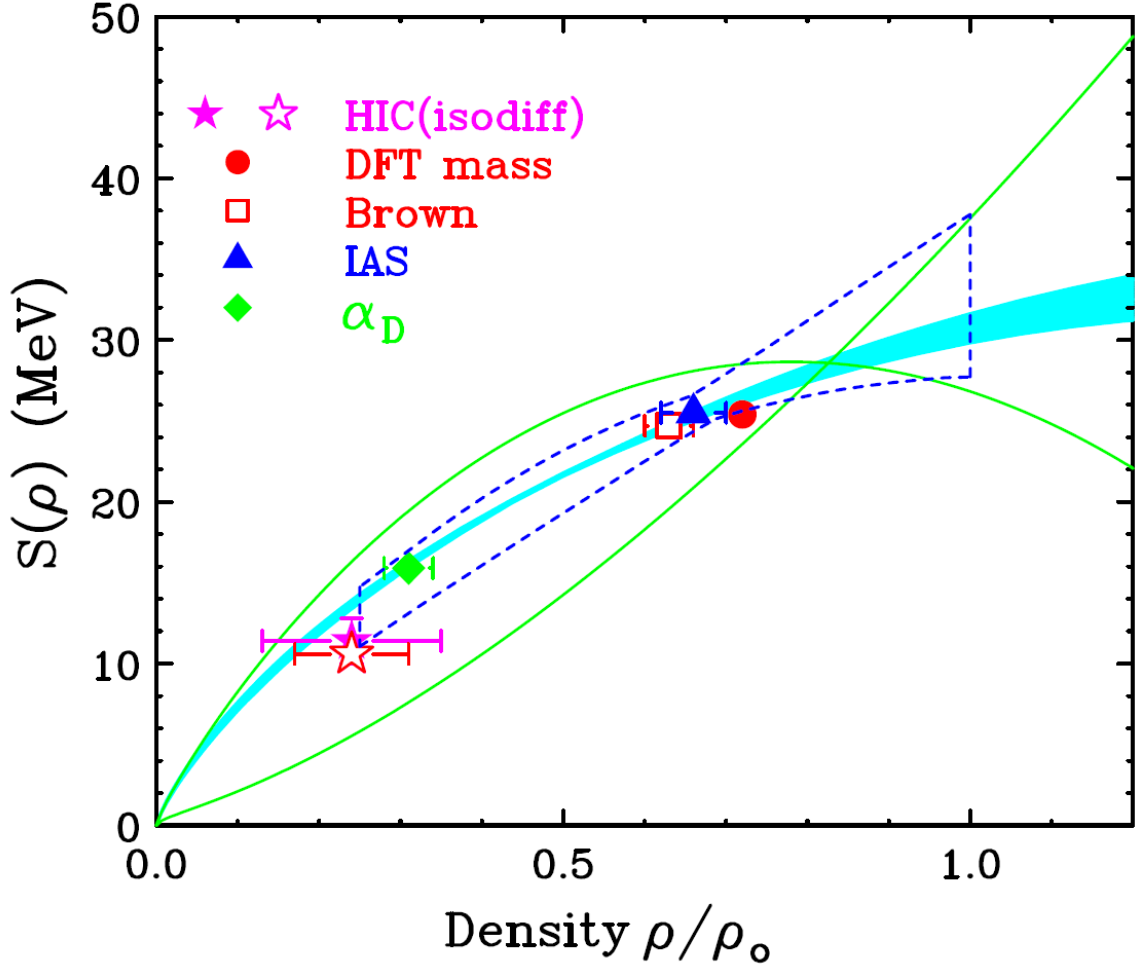

Figure 3: Density dependence of the symmetry energies obtained from the constraint contours in Figure 1. See text for detailed explanation of the symbols and the blue dashed contour. The light blue shaded curve represents the chiral effective field calculations from Ref [32]. The two thin green curves represent SkI1 and Z Skyrme functions [28,38].

The HIC constraints are obtained by modeling the isospin diffusion data of ref. [13] with the transport model ImQMD_2005 code [33] and the symmetry energy function of Eq. (4), which is widely used in transport models [8,33,35].

$$S(\rho) = A(\rho/\rho_0)^{2/3} + B(\rho/\rho_0)^{\gamma}. \tag{4}$$

Isospin diffusion is primarily driven by the symmetry energy at low densities in the neck region between projectile and target nuclei, where it governs the diffusion rate between these nuclei when they have different isospin asymmetries [34]. The slopes of the right (high L) and left (low L) boundaries of the HIC contour in Fig. 1 differ significantly; the analysis via Eqs. 3 & 4 of these slopes consequently leads to a wider range of densities than was obtained for the other observables. Consistent with large slopes of the $2\sigma$ constraint boundaries in Figure 1, the slope analysis provides a lower value for the sensitive density of $\rho_s/\rho_0 = 0.24 \pm 0.11$ and the symmetry energy of $S(\rho_s) = 11.4 \pm 1.4$ MeV, which is plotted as the solid star in Figure 3.

Analogous to Brown's analysis, we can confirm this by directly examining the cross-over of the symmetry energy functions used in ref. [13] to obtain HIC constraint boundary in Figure 1. The curves in the left panel of Figure 4 merges at $S(\rho) \approx 10.6$ MeV and $\rho/\rho_0 \approx 0.17$ and corresponds to symmetry energy functions along the left (low L) $2\sigma$ boundary of the HIC constraint contour of Figure 1. The curves in the right panel of Figure 4 merge at $S(\rho) \approx 10.6 \pm 1.0$ MeV and $\rho/\rho_0 \approx 0.30$ and corresponds to calculations along the right (high L) $2\sigma$ boundary of the HIC constraint contour. This direct examination of the calculations provides a constraint of $S(\rho_s) \approx 10.6 \pm 1.5$ MeV at $\rho_s/\rho_0 = 0.24 \pm 0.7$ shown by the open star in Figure 3, similar to the constraint (solid star) obtained from the slopes of the HIC constraint contour of Figure 1. Thus, this observable probes the lower densities relevant to understanding the EoS of the neutrino-sphere in a core-collapse supernova [4].

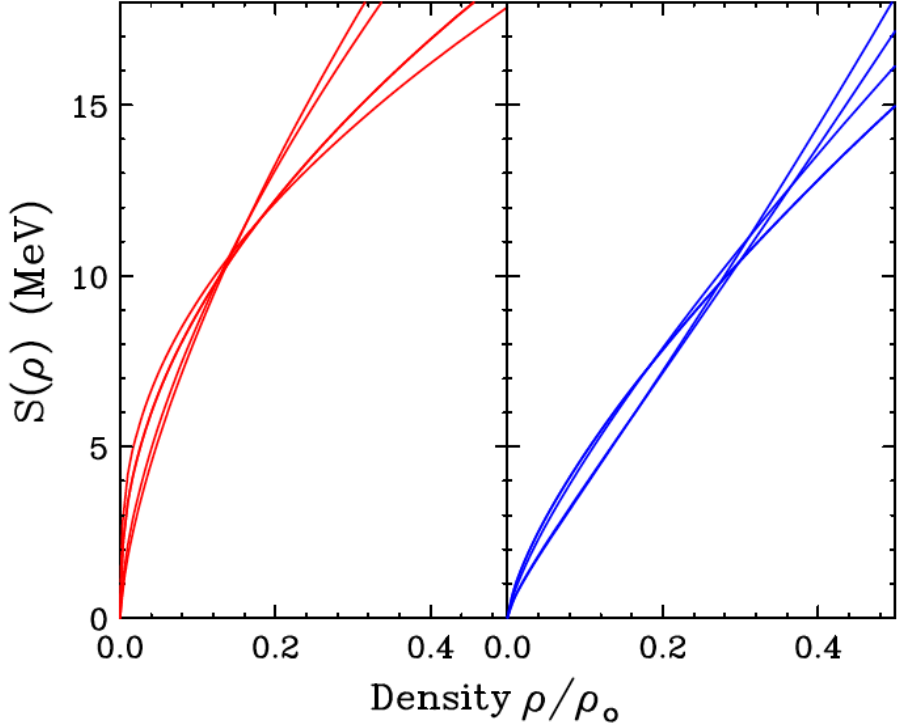

Figure 4: Symmetry energy functions of Eq. (4) corresponding to the 2σ limits from isospin diffusion. Left panel: Symmetry energy functions from the lower (HIC) bounds on L from Figure 1. Right panel: Symmetry energy functions from the upper (HIC) bounds on L from Figure 1.

The $\alpha_D$ contour in Figure 1 was obtained from the electric dipole polarizability of $^{208}$Pb [19,20, 23, 36]. Unlike the mass and IAS observables, however, the polarizability $\alpha_D$ does not represent a simple expectation value of the Hamiltonian but rather the shift in ground state energy due to the presence of an external electric field. Thus the slopes of the observed $\alpha_D$ contours cannot be analyzed with Eq. 3 to extract $S(\rho_s)$ in the same manner as for the (DFT) or (IAS) contours. Difficulty in extracting the sensitive density from constraint contours was noted in ref. [19] where the $\alpha_D$ values for $^{208}$Pb, $^{120}$Sn and $^{68}$Ni were extracted and compared. An alternative direct analysis of the density dependence of the $\alpha_D$ constraint by Zhang et al. [37] shows that $\alpha_D$ probes a range of densities; at the most sensitive density of $\rho_s/\rho_0 \approx 0.31$, ref. [37] extracts $S(\rho_s) = 15.9 \pm 1.0\ MeV$. This point is plotted as the solid diamond (green) in Figure 3.

In the past decade, there have been significant efforts to calculate the equation of state of neutron matter [32] focused on providing predictions of the properties of neutron stars and neutron skins. On the basis of the constraints shown in Figure 3, and the observed deformability of neutron stars, one can exclude 208 out of the 240 Skyrme symmetry energy functions in ref. [28]. As illustrative examples, we show, as thin (green) solid lines in Figure 3, the strongly (SkI1) and weakly

(Z) density dependent Skyrme functions that that have successfully described masses [28,38] but lie well away from the present constraints.

The light blue band in Figure 3 shows ab-initio predictions from Chiral Effective Field Theory (CEFT) that includes 2N and 3N forces [32]. The predicted trend from CEFT compares well at $0.25\rho_0 < \rho < 0.75\rho_0$ to the symmetry energies extracted in this work and to the density dependent constraints obtained from isobaric analog states. Constraint analyses with other forms of the effective interactions, such as relativistic mean field theory and the Gogny force would also be very interesting. Additional constraints on $S(\rho)$ at low densities ($\rho < 0.5\rho_0$) would also help to constrain the curvature of the symmetry energy, i.e. $K_{sym}$ in Eq. (1).

In summary, we have shown how the slopes of $S_0 - L$ correlation can be analyzed to obtain the symmetry energy at a sensitive density $\rho_s$ at which calculations of that observable are most sensitive. We obtain constraints on the symmetry energy over densities of $0.25 \leq \rho/\rho_0 \leq 0.75$ that are relevant to inner crust of neutron stars, to their crustal vibrations and to the neutrino-sphere of core-collapse supernovae. By focusing on the sensitive density, we avoid the uncertainties injected into published $S_0$ and $L$ values during model dependent extrapolations of the symmetry energy from $\rho_s$ to $\rho_0$. When our results are compared to other analyses that extract the symmetry energy at specific densities, consistent constraints emerge. These constraints exclude a variety of symmetry energy terms of the nuclear EoS, especially those with very strong or very weak density dependencies.

**Acknowledgement**:
The authors would like to thank Alex Brown for providing the parameters of the 18 best fit Skyrme functions used in ref [21, 28], Ingo Tews, Jorge Piekarewicz and Pawel Danielewicz for fruitful discussions. This work is supported by the US National Science Foundation Grant No. PHY-1565546, U.S. Department of Energy (Office of Science) under Grant Nos. DE-SC0014530, DE-NA0002923.

**References:**
[1] J.M. Lattimer and M. Prakash, Science 304, 536 (2004).
[2] J.M. Lattimer and M. Prakash, Phys. Rep. 621, 127 (2016).
[3] A.W. Steiner J.M. Lattimer, E.F. Brown, Astrophys. J. 722, 33 (2010).
[4] L. F. Roberts, Sanjay Reddy, and Gang Shen, Phys. Rev. C 86, 065803 (2012).
[5] B. P. Abbott *et al.*, Phys. Rev. Lett. **119**, 161101 (2017).
[6] P. Danielewicz, R. Lacey, and W.G. Lynch Science 298, 1592 (2002).
[7] C.J. Horowitz, E.F. Brown, Y. Kim, W.G. Lynch, R. Michaels, A. Ono, J. Piekarewicz, M.B. Tsang, and H.H. Wolter, J. Phys. G 41, 093001 (2014).
[8] B-A Li, L-W Chen, C.M. Ko, Phys. Rep. 464, 113 (2008).

[9] C.J. Horowitz and A. Schwenk, Nucl. Phys. A 776, 55 (2006).
[10] G. Martinez-Pinedo, T. Fischer, A. Lohs, and L. Huther, Phys. Rev. Lett. 109, 251104 (2012)
[11] K. Hebeler, J. M. Lattimer, C. J. Pethick and A. Schwenk, Astrophys. J. 773, 11 (2013).
[12] Andrew W. Steiner and Anna L. Watts, Phys. Rev. Lett. 103, 181101 (2009).
[13] M.B. Tsang, Y.X. Zhang, P. Danielewicz, M. Famiano, Z.X. Li, W.G. Lynch and A.W. Steiner, Phys. Rev. Lett. 102, 122701 (2009).
[14] M.B. Tsang, J.R. Stone, F. Camera, P. Danielewicz, S. Gandolfi, K. Hebeler, C.J. Horowitz, J. Lee, W.G. Lynch, Z. Kohley, R. Lemmon, P. Moller, T. Murakami, S. Riordan, X. Roca-Maza, F. Sammarruca, A.W. Steiner, I. Vidana, S.J. Yennello, Phys. Rev. C. 86, 015803 (2012).
[15] James M. Lattimer and Yeunhwan Lim, Astrophys. J. 771, 168 (2013).
[16] B.A. Li and X. Han, Phys. Lett. B727, 276 (2013).
[17] M. Kortelainen, T. Lesinski, J. Moré, W. Nazarewicz, J. Sarich, N. Schunck, M. V. Stoitsov, and S. Wild, Phys. Rev. C 82, 024313 (2010).
[18] P. Danielewicz, and J. Lee, Nucl. Phys. A 922, 1 (2014).
[19] X. Roca-Maza, X. Viñas, M. Centelles, B. K. Agrawal, G. Colò, N. Paar, J. Piekarewicz, and D. Vretenar, Phys. Rev. C 92, 064304 (2015).
[20] A. Tamii, I. Poltoratska, P. vonNeumann-Cosel, Y. Fujita, T. Adachi, C. A. Bertulani, J. Carter, M. Dozono, H. Fujita, K. Fujita, K. Hatanaka, D. Ishikawa, M. Itoh, T. Kawabata, Y. Kalmykov, A. M. Krumbholz, E. Litvinova, H. Matsubara, K. Nakanishi, R. Neveling, H. Okamura, H. J. Ong, B. Ozel-Tashenov, V. Y. Ponomarev, A. Richter, B. Rubio, H. Sakaguchi, Y. Sakemi, Y. Sasamoto, Y. Shimbara, Y. Shimizu, F. D. Smit, T. Suzuki, Y. Tameshige, J. Wambach, R. Yamada, M. Yosoi, and J. Zenihiro, Phys. Rev. Lett. 107, 062502 (2011).
[21] B.A. Brown, Phys. Rev. Lett. 111, 232502 (2013).
[22] For clarity, we exchanged the $L$ and $S_0$ axes in Figure 1 with respect to the corresponding figure in ref. [7].
[23] X. Roca-Maza, M. Brenna, and G. Colò, M. Centelles and X. Viñas, B. K. Agrawal, N. Paar and D. Vretenar, J. Piekarewicz, Phys. Rev. C 88, 024316 (2013).
[24] M. B. Tsang, T. X. Liu, L. Shi, P. Danielewicz, C. K. Gelbke, X. D. Liu, W. G. Lynch, W. P. Tan, G. Verde, A. Wagner, H. S. Xu, W. A. Friedman, L. Beaulieu, B. Davin, R. T. de Souza, Y. Larochelle, T. Lefort, R. Yanez, V. E. Viola Jr., R. J. Charity, and L. G. Sobotka, Phys. Rev. Lett. 92, 062701 (2004).
[25] T. X. Liu, W. G. Lynch, M. B. Tsang, X. D. Liu, R. Shomin, W. P. Tan, G. Verde, A. Wagner, H. F. Xi, H. S. Xu, B. Davin, Y. Larochelle, R. T. de Souza, R. J. Charity, and L. G. Sobotka, Phys. Rev. C 76, 034603 (2007).
[26] C. J. Horowitz and J. Piekarewicz, Phys. Rev. Lett. 86. 5647 (2001).
[27] M. Centelles, X. Roca-Maza, X. Viñas, M. Warda, Phys. Rev. Lett. 102, 122502 (2009).
[28] M. Dutra, O. Lourenço, J. S. Sá Martins, A. Delfino, J. R. Stone, and P. D. Stevenson, Phys. Rev. C 85, 035201 (2012).
[29] B. A. Brown and A. Schwenk, Phys. Rev. C 89, 011307(R) (2014).
[30] M. Kortelainen, J. McDonnell, W. Nazarewicz, P.-G. Reinhard, J. Sarich, N. Schunck, M. V. Stoitsov, and S. M. Wild, Phys. Rev. C 85, 024304 (2012).
[31] Paweł Danielewicz, Pardeep Singh, Jenny Lee, Nucl. Phys. A 958, 147 (2017).
[32] C. Drischler, V. Soma, and A. Schwenk, Phys. Rev. C 89, 025806 (2014).
[33] Y. Zhang, M.B. Tsang, ZhuXia Li, Hang Liu, Phys. Lett. B 664, 145 (2008).
[34] L. Shi and P. Danielewicz, Phys. Rev. C 68, 064604 (2003).
[35] J. Hong and P. Danielewicz, Phys. Rev. C **90**, 024605 (2014).
[36] A. Tamii, P. von Neumann-Cosel, and I. Poltoratska, Eur. Phys. J. A 50, 28 (2014).

[37] Zhen Zhang and Lie-Wen Chen, Phys. Rev. C 92, 031301(R) (2015).
[38] B.A. Brown, Phys. Rev. Lett. 85, 5296 (2000).
[39] P. Morfouace et al., private communications.